# Synthesis of nanocrystalline (Co,Ni)Al$_2$O$_4$ spinel powder

# by mechanical milling of quasicrystalline material

**T.P. Yadav[1], N.K. Mukhopadhyay [2], R.S.Tiwari [1] and O.N.Srivastava [1*]**

[1]Department of Physics, Banaras Hindu University, Varanasi-221005, India

[2]Department of Metallurgical Engg., Institute of Technology, Banaras Hindu University, Varanasi 221 005, India

*Author to whom correspondence should be addressed

E-mail : hepons@yahoo.com(ONS),yadavtp@rediffmail.com(TPY)

## ABSTRACT

In the present study, attempts have been made to synthesize the nano-crystalline $(Co,Ni)Al_2O_4$ spinel powders by ball milling and subsequent annealing. An alloy of $Al_{70}Co_{15}Ni_{15}$, exhibiting the formation of a complex intermetallic compound known as decagonal quasicrystal is selected as the starting material for mechanical milling. It is interesting to note that this alloy is close to the stoichiometry of aluminum and transition metal atoms required to form the aluminate spinel. The milling was carried out in an attritor mill at 400 rpm for 40 hours with ball to powder ratio of 20:1 in hexane medium. Subsequent to this annealing was performed in an air ambience for 10, 20 and 40 h at 600 °C in side the furnace in order to oxidize the decagonal phase and finally to form the spinel structure. The X-ray diffraction (XRD) and transmission electron microscopy (TEM) confirmed the formation of nano-sized decagonal phase after milling and then $(Co,Ni)Al_2O_4$ spinel type phase after annealing. The XRD studies reveal the lattice parameter to be 8.075 Å and the lattice strain as 0.6%. The XRD and TEM explorations of spinel phase indicate the average grain size to be ~ 40 nm.

## 1. INTRODUCTION

The $(Co,Ni)Al_2O_4$ is an aluminum transition metal oxide which falls into the category of aluminate spinels. The synthesis of nano-crystalline spinel has been investigated intensively in the recent years due to their unique potential application in high density magnetic recording, microwave devices, magnetic fluids, and also as an absorbent material to remove sulfur-containing materials from hot-coal gas[1,2]. Most of the ceramic dyeing materials used in pigment industries are of transition metal oxides with the spinel structure having high surface area, thermal stability and chemical resistance[3]. The spinel structure is characterized by a simultaneous occupation of tetrahedral and octahedral positions by metallic cations in an FCC oxygen lattice[4]. Cobalt and nickel aluminates have received attention as a ceramic pigment because of its technological significance[5] as well as heterogeneous catalytic effects[6-7]. It has also been proposed as a promising candidate for anode in aluminum production because of its higher resistance to alkalis [8]. The spinel materials can be prepared by many methods such as solid-state reaction[9], hydrothermal[10], alkoxide hydrolysis[11], sol-gel method[12] and microwave-induced method[13]. The most general method is the solid-state reaction, which involves the mixture of metal oxides followed by sintering at high temperature. In the present investigation, we have achieved the formation of $(Co,Ni)Al_2O_4$ spinel phase through the Al-Co-Ni quasicrystalline (qc) decagonal variant .It may be pointed out that qc phases in Al-Co-Ni system are stable and in general quasicrystalline materials are shown to be brittle[14-15]. Therefore, they can be easily converted to the nano particle form through ball milling. The milled powder can be converted to nanophase spinel powder with a moderate temperature annealing in presence of air. Thus this process will have the advantage over the other known process for obtaining the nanophase spinel powder. The as prepared nano-spinel phase may be directly used for many catalytic applications.

In this investigation, we have used high energy ball milling technique which has now become one of the important techniques for the production of metastable nano crystalline/quasicrystalline and amorphous phases[16]. It is necessary to characterize the grain sizes and the defect states of the ball-milled material in order to understand the kinetics of phase transformation. Since, the physical properties of the material depend upon the microstructures and therefore, by controlling the microstructure, materials can be designed according to desirable properties. It can also provide mechanical activation, which enhances the reactivity of the milled powders in subsequent steps[17]. Mechanical activation facilitates the ignition of self-propagation reactions leading to the synthesis of materials in relatively short time. Example of this synthesis technique usually referred as mechanically activated self-propagation (MASP) have been reported recently[18].

In this paper, we have shown that due to the brittleness and closeness of the stoichiometry towards spinel phase, the decagonal quasicrystalline $Al_{70}Co_{15}Ni_{15}$ powder plays an important role for the formation of $(Co,Ni)Al_2O_4$ spinel. To the best of our knowledge, so far the $(Co,Ni)Al_2O_4$ spinel structure has not yet been synthesized from quasicrystalline alloy by ball milling and annealing. In the present paper we report the preparation and characterization of $(Co,Ni)Al_2O_4$ nano spinel powder obtained through $Al_{70}Co_{15}Ni_{15}$ decagonal alloy by ball milling and controlled annealing in air.

## 2. EXPERIMENTAL PROCEDURE

### a) Synthesis

Raw materials with a composition close to $Al_{70}Co_{15}Ni_{15}$ (Al=99.98%, Co=99.99%, Ni=99.97% pure) were selected for pelletization. The pallet was melted in an R.F. induction furnace, in the presence of dry argon atmosphere. The ingot formed was re-melted several times to ensure homogeneity. The as-cast ingot was crushed to particles less than 0.5 mm in size and placed in an attritor ball mill (Szegvari Attritor) with a ball to powder weight ratio 20:1. The attritor has a cylindrical stainless steel tank of inner diameter 13 cm and the angular speed of mill was maintained at 400 rpm, which is the maximum speed that can be attained with the present experimental set up. It can be mentioned that at lower speed when the milling intensity is expected to be low, the single nano-decagonal phase was not obtained within the given maximum milling time. Therefore, after several trials, it was found that the milling intenstity corresponding to speed of 400 rpm is best for our experiment to obtain nano-decagonal phase. As far as the process controlling agent is concerned, the hexane medium produced the optimum results in regards to nano-decagonal phase formation. The milling operation was conducted from 5 to 40 h. The powder obtained after 40h of milling was annealed at 600 $^{o}$C for time ranging from 10 to 40h in air ambience.

**b) Characterizations**

The milled and heat-treated powders were characterized by powder X-ray diffraction (XRD) using a Philips 1710 X-ray diffractometer with CuK$\alpha$ radiation ($\lambda$=1.54026Å). The effective crystallite size and relative strain of mechanically milled / mechanical activated powders as well as heat-treated products were calculated based on line broadening of XRD peaks. The use of the Voigt function for the analysis of the integral breadths of broadened X-ray diffraction line profiles forms the basis of a rapid and powerful single line method of crystallite-size and strain determination. In this case the constituent Cauchy and Gaussian components can be obtained from the ratio of full width at half maximum intensity (2ω) and integral breadth. In a single line analysis     the apparent crystallite size ' D ` and strain 'e ` can be related to Cauchy ($\beta_c$) and Gaussian ($\beta_G$) widths of the diffraction peak at the Bragg angle $\theta$[19]. The transmission electron microscopy using Philips CM-12 has been used for micostructural and structural characterization with operating voltage at 100 kV. An energy-dispersive X-ray (EDX) link with a transmission electron microscope (TEM), Tecnai 20 G$^2$ system was employed for the compositional analysis. The magnetic property, magnetic hysterisis measurements were carried out using a vibrating sample magnetometer by applying maximum field of 10 kG at room temperature

## 3. RESULTS AND DISCUSSIONS

Fig. 1(a) shows the XRD pattern of 40 h mechanical milled (MM) $Al_{70}Co_{15}Ni_{15}$ powder exhibiting the formation of nano- decagonal phase. The decagonal phase diffraction pattern was indexed by using six independent indices[19]. All of the peaks are indexed based on decagonal phase. No other phase could be identified. It should be mentioned that in Al-Co-Ni system, the existence of the thermodynamically stable and the perfect decagonal quasicrystal has already been reported in the literature[20]. The XRD pattern of 40h MM powders exhibits the broadening of peaks, which can be attributed to the decrease in the particle size and also an increase in internal strain. The different stages of annealing for 10h, 20h and 40h in air at 600 $^o$C show the gradual formation of $(Co,Ni)Al_2O_4$ type spinel (Fig. 1(b-d)). The XRD pattern (Fig. 1d) shows the peak corresponding to the spinel structure with the lattice parameter (FCC, a=8.075±0.004A). The calculated density was found to be 4.45g/cc, where as the measured density was found to be 4.39-g/cc confirming the formation of mixed spinel structure containing Ni and Co cations randomly in the tetrahedral and octahedral sites in structures. From these results, it is obvious that high rate of oxygen diffusion through the nano-decagonal phases has led to the formation of spinel structures. It may be mentioned that heat treatment of the as-cast decagonal phase at 600 $^{o}$C temperature was also carried out and it was found that the formation of spinel was not possible within the time of 40 h. Only a little amount of aluminum oxide was found to be formed. Hence the quasicrystal phase alone was not advantageous to form the spinel phase. It appears that the nanodecagonal phase is prerequisite for the formation of spinel phase under the above mentioned conditions of annealing temperature and duration. This can be understood that nanodecagonal phase obtained during the MM, has increased the diffusivity by creating the short diffusion path by increasing the surface area due to nanophase microstructures and also by increasing the surface activity due to high concentration of defects. Thus it is expected that a considerable reduction in the activation energy for the diffusivity will result. Hence, the kinetics of oxidation rate, which is controlled by oxygen diffusion, becomes faster compared to the micro-quasicrystalline phase, which is present in as-solidified quasicrystalline material. This is the reason that the spinel formation was not possible in case of miscro-quasicrystalline material even after providing the same annealing treatment. From the analysis of the crystallite sizes and the strains it is found that after the 40h MM the average crystallite size is ~14nm and lattice strain is ~0.98 %. During annealing final crystallite size increases and goes up to 40 nm but still it is within nanometer scale. This also implies that during the phase formation, any significant grain growth of the spinel has not taken place. This is because of the fact that the phase transformation, which is caused by the conversion of decagonal variant into spinel phase, inhibits the significant grain coarsening.

The bright field and dark-field microstructures of 40 h MM and subsequent annealed sample (40 h annealing in air at 600 ° C) are shown figs. 2(a) and (b). The micro-twinning   of the particle during transformation of decagonal phase to spinel phase has been marked by arrow.  The average particle size appears to be consistent with data obtained from XRD. It is clearly visible that fine grains are having size ranging from 40 to 60nm. It may be pointed out that the existence of a wide range of sizes in ball-milled and annealed sample has been previously reported[21]. The corresponding selected area diffraction pattern identifies the presence of spotty rings due to the    $(Co,Ni)Al_2O_4$ spinel phase with lattice parameter of 8.075Å  of  FCC.

The compositional analysis of 40h MM and air-annealed powder was analyzed by energy depressive X-ray analysis (EDX) attached to a transmission electron microscope (Fig. 3). Some contamination of Fe from stainless steel balls was observed (Fig. 3(a)). It also shows the presence of oxygen approximated to be around 6 wt% in as-milled sample. However, after annealing in air the oxygen wt% increases up to 64 wt% indicating the oxygen pick up from air during annealing. It should be noted that the oxygen pick up in case of micro-quasicrystalline sample was not significant compared to that of MM powder when both the samples were annealed in air. The contamination of Fe appears to be not more than 0.19 wt% in MM powder, which is not expected to cause any phase transformation as per the stability of the decagonal phase.

The magnetic properties of the nano-decagonal and nano $(Co,Ni)Al_2O_4$ spinel have been observed by simple M-H curve (Fig. 4a,b). The 40h milled $Al_{70}Co_{15}Ni_{15}$ powder shows the soft ferromagnetic behaviors with $M_s$ = 0.28483 and $H_c$ = 0.4065 KG. Whereas the corresponding values for spinels are found to be $M_r$ =3.1673 and $H_c$ =1.20625 KG. The quasicrystalline samples are known to exhibit paramagnetic behaviour. However, due to the formation of nanoquasicrystalline phases along with some contamination by Fe from milling media, the weak ferromagnetic behaviour (close to superparamagnetic behaviour) has been observed. This behaviour is itself unique and needs further investigation. After annealing the ferromagnetic behaviour is much more prominent as the $M_s$ and $M_r$ and $H_c$ have changed significantly. The magnetic properties of the spinel materials are enhanced quite significantly compared to the milled nanodecagonal phase, which can be attributed to its crystal structure. Generally the inverse spinel structure shows ferromagnetic behaviour. It is interesting to point out that the nickel aluminate spinal $NiAl_2O_4$ is almost inverse spinel with the nickel ion preferentially distributed over the octahedral site [21] but the cobalt aluminate spinels are generally normal spinels in microscales. The effect of nanocrystallinity, inverse spinel structure and the Fe contamination from milling media has overall effects on the magnetic properties of the present spinel material. In case of nanospinels, the ferromagnetic properties are distinct although it is lower than the ferrite spinels, as the structure here is not fully inverse spinel type. It can be emphasized that a small amount Fe which is picked up during milling will not be able to affect the magnetic property so drastically in the case of the spinel phase. More investigation is required to understand the effect of annealing time and temperature on the formation of spinel phase, and the Curie temperature. It is also necessary to establish the detailed distributions of cations in various sites and their effect on ferromagnetic behaviour. However, the implication of the present work is to emphasize the possibility of a new route for the synthesis of nanospinel structures along with the development of ferromagnetic properties. At this stage micromechanism of phase transformation of nano-decagonal variant to nano-spinel phase appears to be faster, though the exact structural rearrangement and the orientation relationships between the initial and final phases has yet to be worked out.

The main objective of the present work has been to synthesise spinel phase through mechanical milling and annealing in air ambience at moderate temperature. The mechanical milling of the quasicrystalline phases is easier because of its brittle nature. Thus the resultant nanoquasicrystalline phases with enhanced reactivity give rise to the spinel phase formation during the calcinations at moderate temperature. In general, the calcinations temperature is more than 1500 $^o$C for the formation of spinels from the mixed oxides by the conventional routes. Moreover, the nanospinel phase formation will be difficult to obtain at this temperature. Since in many technological applications nanospinel powders are used a catalyst due to its high reactivity and thermal stability, the present route of spinel formation, which gives rise to nano-powder phase of spinel is of significant importance [22].

## 4. CONCLUSIONS

In the present study, the synthesis and magnetic behaviour of $(Co,Ni)Al_2O_4$ spinel through milling of decagonal phase and subsequent annealing in air have been investigated. It has been found, that after 40 h ball milling of $Al_{70}Co_{15}Ni_{15}$ alloys, nanosized decagonal phase evolved, which after subsequent annealing (40 h at 600 °C in air) transformed to nano $(Co,Ni)Al_2O_4$ spinel phase.. Thus the synthesis of nanostructured $(Co,Ni)Al_2O_4$ spinel powders at moderate temperatute by this route appears to be an interestingly new route for the production of nano spinels. Significant enhancement in the magnetic properties has been found in nano-spinels compared to that of nano-decagonal phase, which can be attributed to the spinel structures. The weak ferromagnetic close to superparamagnetic behaviour is established for the first time in the nano-quasicrystalline phase. In case of nanospinels, the ferromagnetic properties are distinct although it is lower than the ferrite spinels.

**Acknowledgement:**

The authors would like to thanks Prof. G.V.S Sastry, Prof D.H.Kim, Dr. Bikramjit Basu and Dr. G.D. Varma for many stimulating discussions. The financial support from the Department of Science & technology DST (NSTI), New Delhi, India is gratefully acknowledged. One of the authors (TPY) acknowledges the support of CSIR for a senior research fellowship.

**References:**

1. S. Bid and S.K. Pradan, Materials Chemistry and Physics  82 27 **(2003).**

2. R.E. Ayala and   D.W. Marsh, Ind. Chem. Res. 30 55 **(1991)**

3. N. Ouahdi, S. Guillemet J.J. Demai, B. Durand L. Er. Rakho, R.Moussa and A.

   Samdi,   Material Letters 95  334 **(2005).**

4. M. Zayat, W.D. Levgy, Chem. Mater. 12 263 **(2000)**

5. Woo-Seok Cho and Masato Kakihana, J. Alloys and Comps.,287  87 **(1999).**

6. G. Busca, V. orenzelli, V.S. Escribano and R. Guidetti, J. Catal 131 167 **(1991)**

7. Yong Sheng Han Jian Bao Li, Xiao Shan Ning Xiao Zhan Zang and Bo Chi

   Materials Science Engineering A369  241 **(2004).**

8. L. Kou, J.R. Selman, J. Appl. Electrochem 30  1433 **(2000).**

9. F.S. Pettit, E.H. Randkler, E.J. Fellten, J. Am. Ceram. Sco 49 199 **(1956).**

10. R. Roy, J. Am Ceram Soc. 36  207 **(1956).**

11. F Duboudin J. Dunogues J. Ravez P Hagenmuller Mater Res. Bull 22 261

    **(1987).**

12. P. Jeevanandanm, Yu Koltypin, A. Gredanken Mater Sci Eng. B 90, 125

    **(2002).**

13. R.D. Peelamedu, R.Roy, D.K. Agrawal Mater Lett. 55  234 **(2002)**

14. N.K.Mukhopadhyay, G.C.Weatherly,J.D.Embury, Mater Sci Eng A 315 202

    **(2001)**

15.  A.P.Tsai, A.Inove and T.Masumoto J. mate Trans JIM 30 300 **( 1989)**

16. T.P. Yadav, N.K. Mukhopadhyay,  R.S.Tiwari and O.N.Srivastava, Phil. Mag

    86 397 **(2006).**

17. U. Anselmi-Tamburini, F. Maglia, S. Doppiu, M. Monagneddu, G. Cocco and Z.A. Murir, J. Mater Res. 19 2558 **(2004).**

18. T.P. Yadav, N.K. Mukhopadhyay,  R.S.Tiwari and O.N.Srivastava, Mater. Sci. Eng. A 393  366 **(2005).**

19. N.K. Mukhopadhyay and E.A. Lord Act Cryst A 58   424 **(2002).**

20. T.P. Yadav, N.K. Mukhopadhyay and O.N.Srivastava, J. Noncrystlaline Solids 334 & 335  57 **(2004).**

21. C.O. Arean, M.P. Mentrait, A.J.L. Lopez, J.B. Parra, Colloids Surf. A : Physico   Chem. Engg. Aspects 180   253 **(2001).**

22. Sivaraj Chokkaram, Ram Srinivasan, Diane R. Milburn, Burtron H. Davis Journal of Molecular Catalysis A: Chemical 121 157 **(1997).**

**Figure Captions:**

**Fig.1.**  X-ray diffraction pattern obtained from the powder after mechanical milling

for 40 h ( curve `a ` ) and subsequent annealing at 600˚ C for 10 –40 h

( curve b – d ). Curve d indicates the formation of nanospinel phase.

**Fig.2. (a)** Bright field and (b) dark field TEM images of powder sample after 40 h

of milling and 40 h of annealing , showing the aggregates of nano spinel of the order of 30-50 nm in size, (c) the corresponding diffraction ring pattern indexed on the basis of  the spinel  phase.

**Fig.3.** Energy –dispersive spectrum of the powder milled for 40h ( curve a) and subsequently annealed at 600 ˚ C for 40  h ( curve b). The Fe peak indicates the  contamination  of the sample during MM .

**Fig.4.** (a) The typical hysteresis loop  (*M-H* curve)  of 40 h MM  $Al_{70}$ $Co_{15}$ $Ni_{15}$

powder.  (b) The *M-H*  curve of $(Co,Ni)Al_2O_4$  nano spinel.

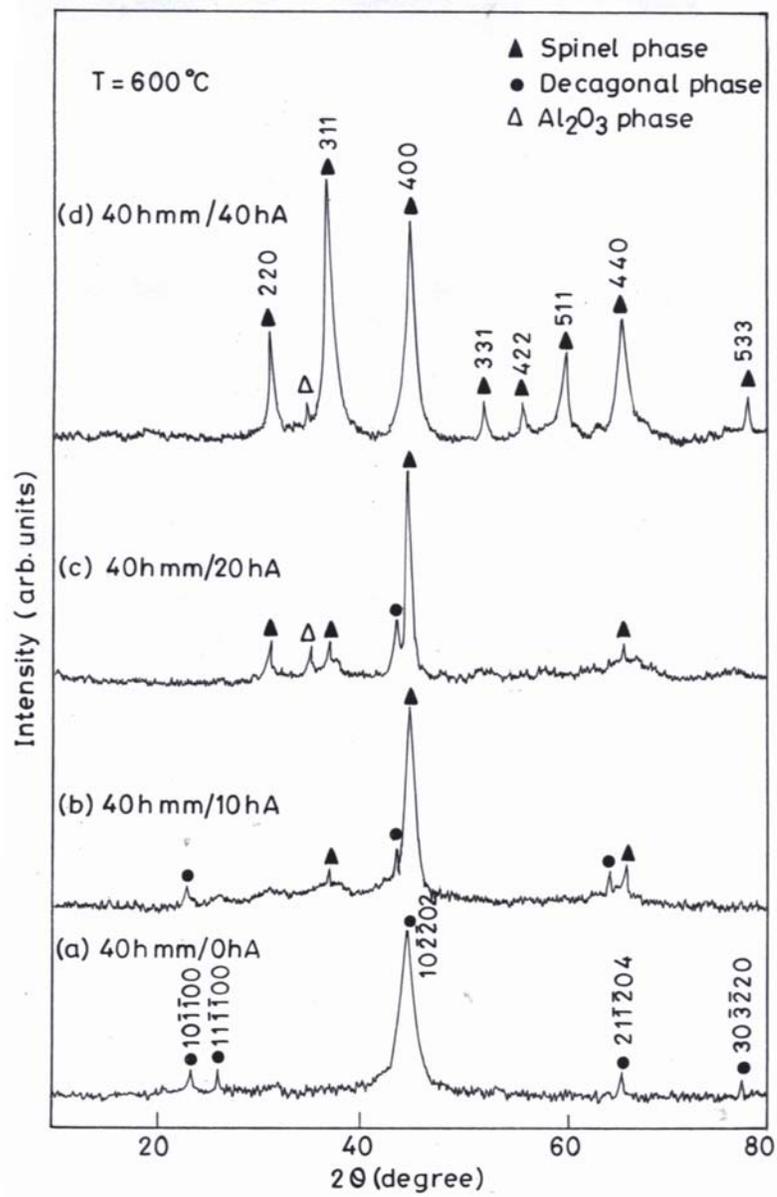

**Fig.1**

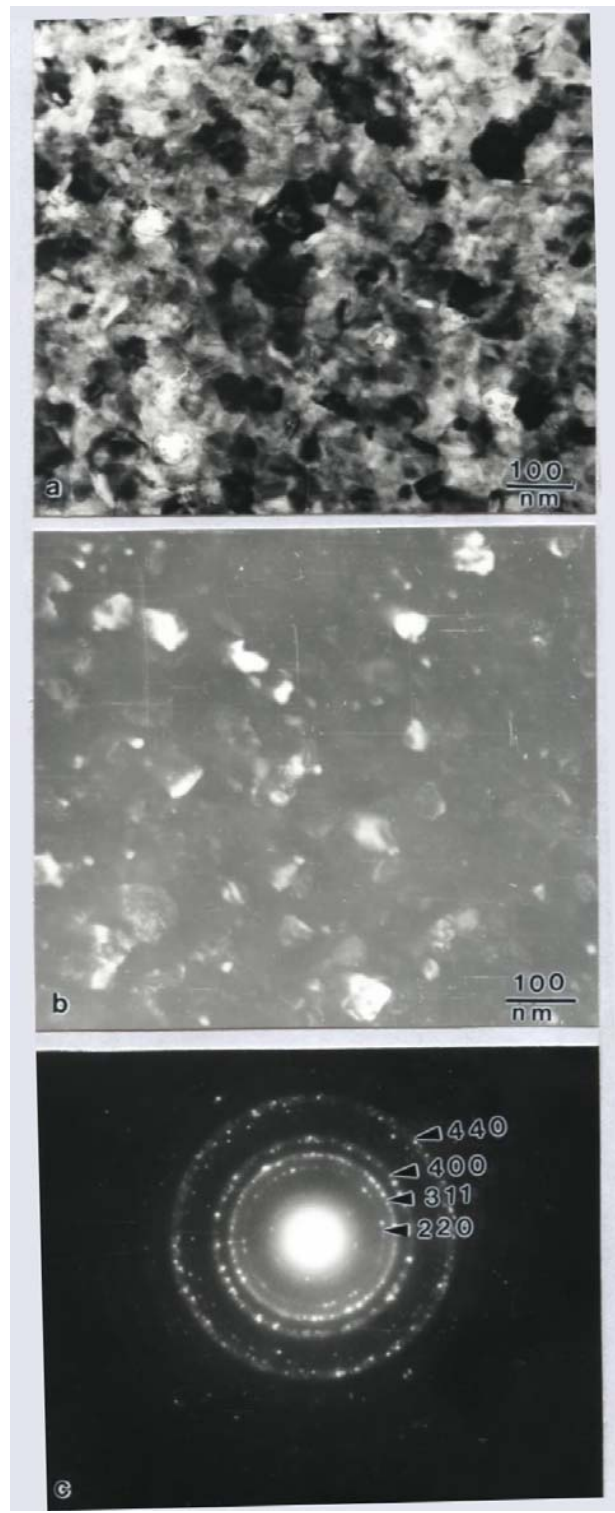

**Fig.2**

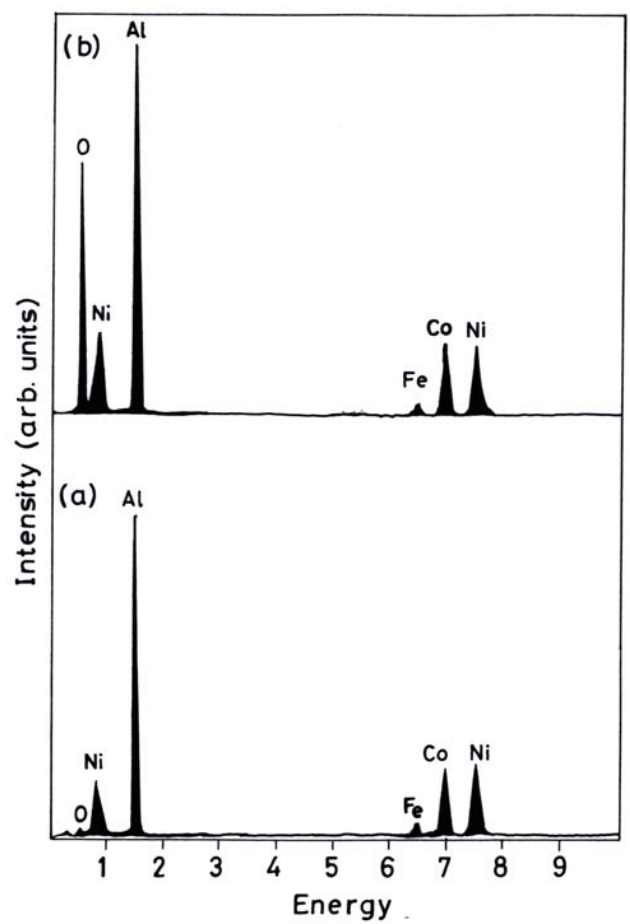

**Fig.3**

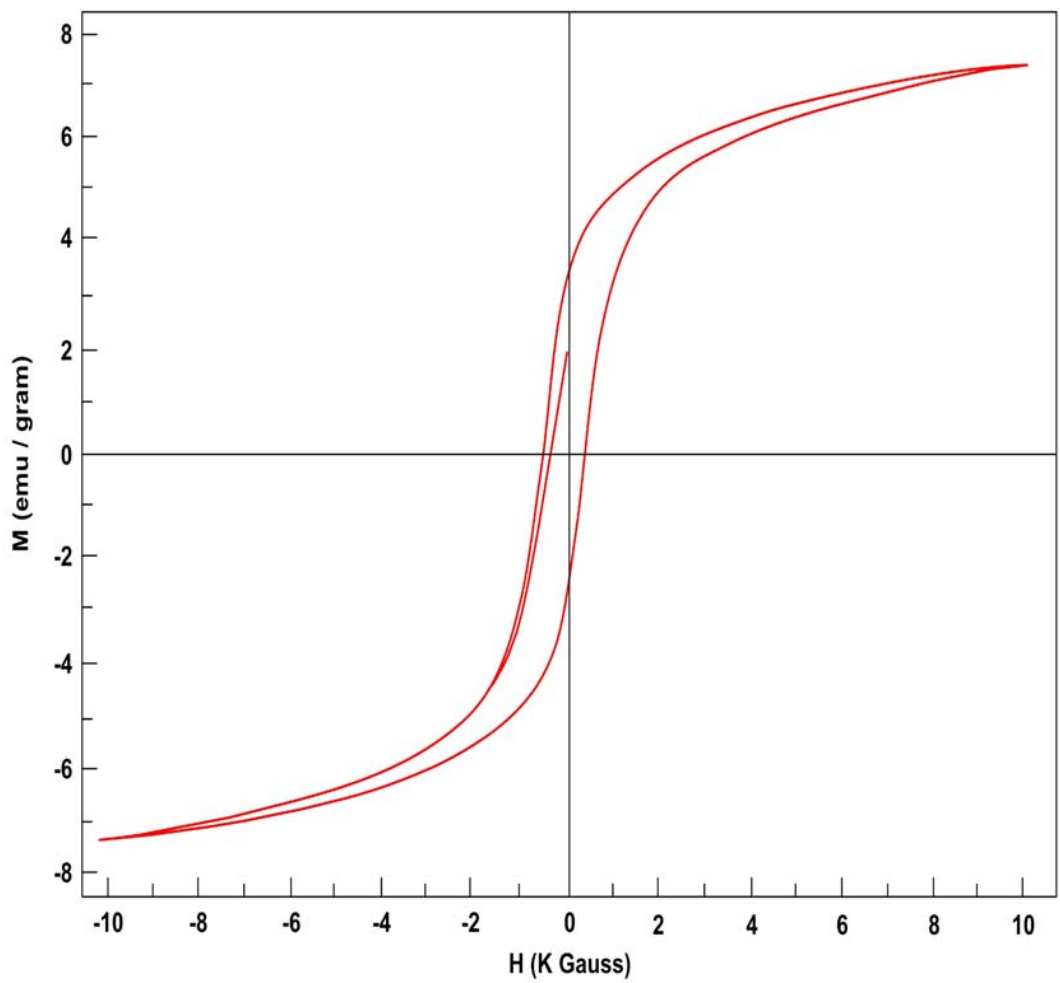

**Fig.4(a)**

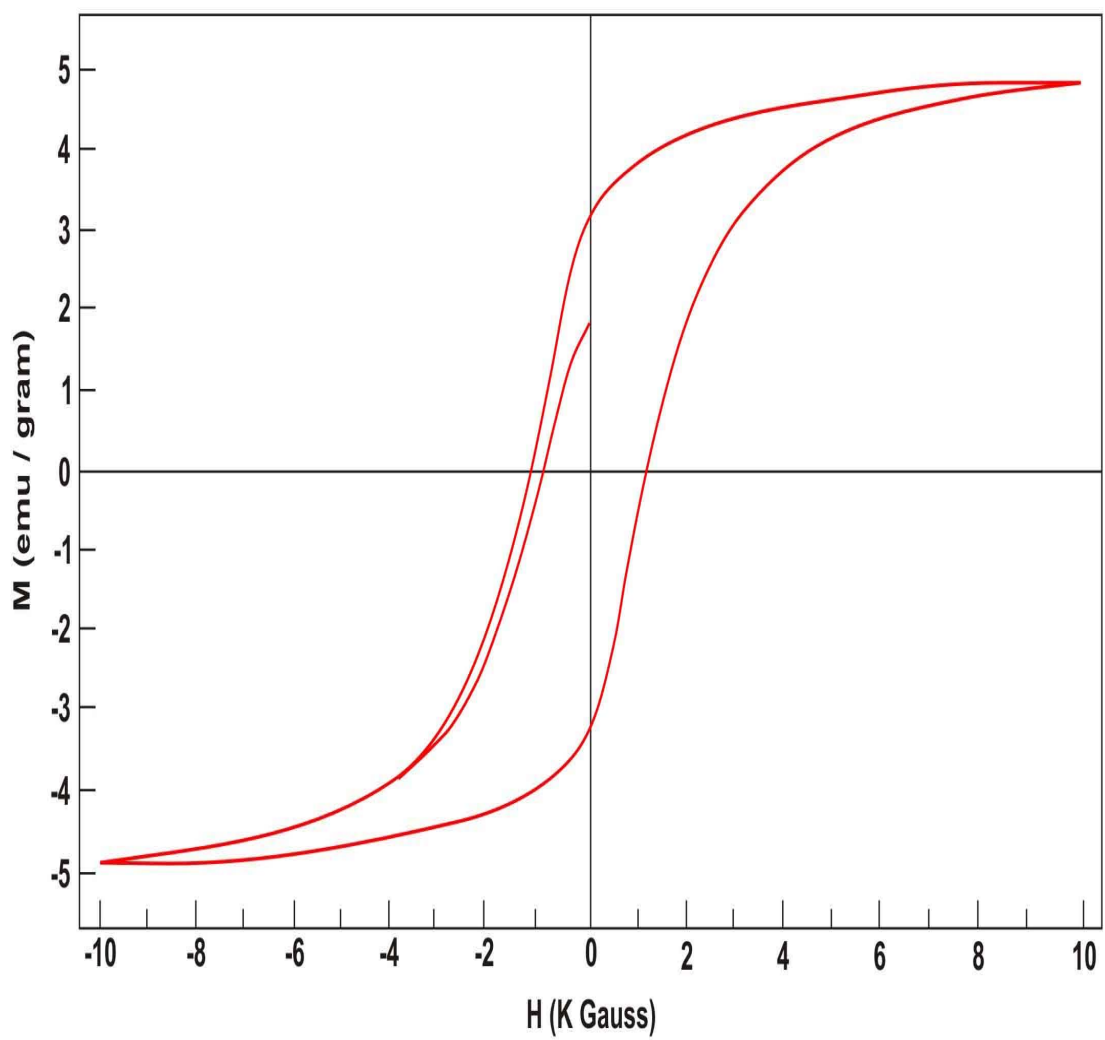

**Fig.4(b)**